\documentclass[aps,prb,showpacs,twocolumn,superscriptaddress]{revtex4}
\usepackage{amsmath,graphics}
\usepackage[next]{inputenc}
\usepackage[dvips]{epsfig}
\def\be{\begin{equation}}
\def\ee{\end{equation}}

\def\bi{\begin{itemize}}
\def\ei{\end{itemize}}
\def\bn{\begin{enumerate}}
\def\en{\end{enumerate}}
\def\bea{\begin{eqnarray}}
\def\eea{\end{eqnarray}}
\def\no{\nonumber}
\def\ba{\begin{array}}
\def\ea{\end{array}}
\def\bd{\begin{displaymath}}
\def\ed{\end{displaymath}}

\begin{document}
\title{Phase Diagram of spin 1/2 XXZ Model With Dzyaloshinskii-Moriya Interaction}

\author{R. Jafari}
\affiliation{Institute for Advanced Studies in Basic Sciences,
Zanjan 45195-1159, Iran}
\affiliation{School of physics, IPM
(Institute for Studies in Theoretical Physics and Mathematics),
\\P. O. Box: 19395-5531, Tehran, Iran}
\author{A. Langari}
\affiliation{Physics Department, Sharif University of Technology,
Tehran 11155-9161, Iran}
\email[]{langari@sharif.edu}
\homepage[]{http://spin.cscm.ir}

\begin{abstract}
We have studied the phase diagram of the one dimensional XXZ model
with Dzyaloshinskii-Moriya (DM) interaction. We have applied the
quantum renormalization group (QRG) approach to get the stable
fixed points, critical point and the scaling of coupling
constants. This model has three phases, ferromagnetic, spin-fluid and
N\'{e}el phases which are separated by a critical line which
depends on the DM coupling constant. We have shown that the staggered
magnetization is the order parameter of the system and investigated
the influence of DM interaction on the chiral ordering as a
helical magnetic order.
\end{abstract}
\date{\today}


\maketitle
%%%%%%%%%%%%%%%%%%%%%%%%%%%%%%%%%%%%%%%%%%%%%%%%%%%%%%%%%%%%%%%%%%%%%
\section{Introduction \label{introduction}}
Quantum phase transition has been one of the most interesting
topic in the area of strongly correlated systems in the last
decade. It is a phase transition at zero temperature where the
quantum fluctuations play the dominant role \cite{vojta}.
Suppression of the thermal fluctuations at zero temperature
introduces the ground state as the representative of the system.
The properties of the ground state may be changed drastically
shown as a non-analytic behavior of a physical quantity by
reaching the quantum critical point (QCP). This can be done by
tuning a parameter in the Hamiltonian, for instance the magnetic
field or the amount of disorder. The ground state of a typical
quantum many body systems consist of a superposition of a huge
number of product states. Understanding this structure is
equivalent to establishing how subsystems are interrelated, which
in turn is what determines many of the relevant properties of the
system. In Mott insulators the Heisenberg interaction is in most
cases the dominant source of coupling between local moments, and
most theoretical investigations are based on modeling in which
only this type of interaction is included. Recently some novel
magnetic properties in antiferromagnetic (AF) systems were discovered
in the variety of quasi-one dimensional materials that are known
to belong to an antisymmetric interaction of the form
$\overrightarrow{D}.(\overrightarrow{S_{i}}\times\overrightarrow{S_{j}})$
which is known as the Dzyaloshinskii-Moriya (DM)interaction. The
relevance of antisymmetric superexchange interactions in spin
Hamiltonian which leads to either a week ferromagnetic (F) or
helical magnetic distortion in quantum AF systems, has been
introduced phenomenologically by
Dzyaloshinskii\cite{Dzyaloshinskii}. A microscopic model of
antisymmetric exchange interaction was first proposed by
Moriya\cite{Moriya} which showed that such interactions arise
naturally in perturbation theory due to the spin-orbit coupling in
magnetic systems with low symmetry and is essentially an extension
of the Anderson superexchange mechanism\cite{Anderson} that shows
for spin-flip hopping of electrons. Since it (DM interaction)
breaks the fundamental SU(2) symmetry of the Heisenberg
interactions, it is at the origin of many deviations from pure
heisenberg behavior such as canting\cite{Coffey} or small
gaps\cite{Dender1,Oshikawa1,Zhao,Fouet,Chernyshev}. A number of AF
systems expected to be described by DM interaction, such as
$Cu(C_{6}D_{5}COO)_{2}3D_{2}O$\cite{Dender1,Dender2},
$Yb_{4}As_{3}$\cite{Kohgi,Fulde,Oshikawa2},
$BaCu_{2}Si_{2}O_{7}$\cite{Tsukada}, $\alpha-Fe_{2}O_{3}$,
$LaMnO_{3}$\cite{Grande} and $K_{2}V_{3}O_{8}$\cite{Greven}, which
exhibit unusual and interesting magnetic properties in the
presence of quantum fluctuations and/or applied magnetic
field\cite{Grande,Yildirim,Katsumata}. Also belonging to the class
of DM antiferromagnets is $La_{2}CuO_{4}$, which is a parent
compound of high-temperature superconductors\cite{Kastner}. This
has stimulated extensive investigation on the physical properties
of the DM interaction. However, This interaction is rather
difficult to Handel analytically, which has brought much
uncertainty in the interpretation of experimental data and has
limited our understanding of many interesting quantum phenomena of
low-dimensional magnetic materials.

In the present paper, we have considered the one dimensional XXZ
model with DM interaction by implementing the quantum
renormalization group (QRG) method. In the next section the QRG
approach will be explained and the renormalization of coupling
constant are obtained. In section (III), we will obtain the phase
diagram, fixed points, critical points and calculate the staggered
magnetization as the order parameter of the underling quantum phase transition. We will also
introduce the chiral order as an ordering which is produced by DM interaction.
The exponent which shows the divergence of correlation function
close to the critical point $(\nu)$, the dynamical exponent $(z)$
and the exponent which shows the vanishing of staggered
magnetization near the critical point $(\beta)$ will also be
calculated. In Sec (IV), discussion concludes the paper.

%%%%%%%%%%%%%%%%%%%%%%%%%%%%%%%%%%%%%%%%%%%%%%%%%%%%%%%%%%%%%%%%%%%%%%%%%%%%%

\section{Quantum renormalization group\label{sec2}}

The main idea of the RG method is the elimination or the thinning
of the degrees of freedom caried out step by step in an iteration procedure.
Here, we used the well known  Kadanoff block method as it is both
well suited to perform analytical calculations in the lattice models and is
conceptually straight-forward to be extended to the higher
dimensions\cite{miguel1,miguel-book,Langari,Jafari}. In the
Kadanoff's method, the lattice is divided into blocks in which the
Hamiltonian can be exactly diagonalized. Selecting a number of
low-lying eigenstates of the blocks the full Hamiltonian is
projected onto these eigenstates and an the effective
(renormalized) Hamiltonian is obtained.

The Hamiltonian of XXZ model with DM interaction in the $z$
direction on a periodic chain of $N$ sites is

%%%%%%%%%%%%%%%%%%%%%  Fig.1   %%%%%%%%%%%%%%%%%%%%%%%
\begin{figure}
\begin{center}
\includegraphics[width=8cm]{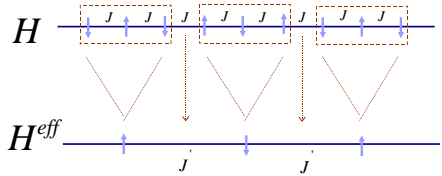}
\caption{(color online)The decomposition of chain into three site
blocks Hamiltonian ($H^{B}$) and inter-block Hamiltonian
($H^{BB}$).} \label{fig1}
\end{center}
\end{figure}
%%%%%%%%%%%%%%%%%%%%%%%%%%%%%%%%%%%%%%%%%%%%%%%%%%%%%%%

\bea \label{eq1}
H(J,\Delta)=\frac{J}{4}\sum_{i}^{N}\Big[\sigma_{i}^{x}\sigma_{i+1}^{x}+\sigma_{i}^{y}\sigma_{i+1}^{y}+
\Delta\sigma_{i}^{z}\sigma_{i+1}^{z}\\
\no
+D(\sigma_{i}^{x}\sigma_{i+1}^{y}-\sigma_{i}^{y}\sigma_{i+1}^{x})\Big],
\eea

where the $J$ is exchange constant, $D$ is the strength of $z$
component of DM interaction and the easy-axis anisotropy defined
by $\Delta$ which can be positive and negative. The positive and
negative $J$ corresponds to the antiferromagnetic and
ferromagnetic (F) cases, respectively. $\sigma_{i}^{\alpha}$
refers to the $\alpha$-component of the Pauli matrix at site $i$.
By implement $\pi$ rotation around $z$ axis on odd or even sites,
the AF case of Hamiltonian ($J>0$) is mapped on the F case ($J<0$) with
opposite sign of anisotropy,

\bea \label{eq5}
H(J,\Delta)=\frac{J}{4}\sum_{i}^{N}\Big[\sigma_{i}^{x}\sigma_{i+1}^{x}+\sigma_{i}^{y}\sigma_{i+1}^{y}-
\Delta\sigma_{i}^{z}\sigma_{i+1}^{z}\\
\no
+D(\sigma_{i}^{x}\sigma_{i+1}^{y}-\sigma_{i}^{y}\sigma_{i+1}^{x})\Big],~~J>0.
\eea

So we can restrict ourselves to AF case ($J>0$) with $D>0$ and
arbitrary anisotropy ($\Delta<0$ and $\Delta>0$) without loss of
generality.

The effective Hamiltonian to the first order RG
approximation is
\bea \no H^{eff}=H^{eff}_{0}+H^{eff}_{1},~~~~~~~~~\\
\no
H^{eff}_{0}=P_{0}H^{B}P_{0}~~~,~~~H^{eff}_{1}=P_{0}H^{BB}P_{0}.
\eea

We consider a three-site block procedure defined in
Fig.(\ref{fig1}). The block Hamiltonian $(H_{B}=\sum h_{I}^{B})$,
its eigenstates and eigenvalues are given
in Appendix A. The three-site block Hamiltonian has four doubly
degenerate eigenvalues (see Appendix A). $P_{0}$ is the projection
operator of the ground state subspace defined by
$\big(P_{0}=|\Uparrow\rangle\langle\psi_{0}|+|\Downarrow\rangle\langle\psi_{0}'|\big)$,
Where $|\psi_{0}\rangle$ and $|\psi_{0}'\rangle$ are the doubly
degenerate ground states, $|\Uparrow\rangle$ and
$|\Downarrow\rangle$ are the renamed base kets in the effective
Hilbert space. For each block we keep two states ($|\psi_{0}\rangle$ and
$|\psi_{0}'\rangle$) to define the effective (new)
site. Thus, the effective site can be considered as having a spin 1/2.
Due to the level crossing which occurs for the
eigenstates of the block Hamiltonian, the projection operator ($P_{0}$) can be
different depending on the coupling constants. Therefore, we must
specify different regions with the corresponding ground states. As  Fig.(\ref{fig5}) shows
there are two regions with different eigenstates which are separated by $\Delta<-\sqrt{1+D^{2}}$
where a level crossing occurs.
In region (A) the ground state is the doubly-degenerate ferromagnetic state $|\psi_{3}\rangle$
and $|\psi_{3}'\rangle$ while in region (B) $|\psi_{0}\rangle$ and $|\psi_{0}'\rangle$ are
the degenerate ground states. At the level crossing  ($\Delta=-\sqrt{1+D^{2}}$) the ground state is 4-fold degenerate ($|\psi_{3}\rangle$, $|\psi_{3}'\rangle$, $|\psi_{0}\rangle$, $|\psi_{0}'\rangle$). A summary of this information is given in Fig.(\ref{fig5}) of appendix A.

In the following, we will classify the RG equation of the regions where each of this states represent the ground
state.

%%%%%%%%%%%%%%%%%%%%%  Fig.2   %%%%%%%%%%%%%%%%%%%%%%%
\begin{figure}
\begin{center}
\includegraphics[width=8cm]{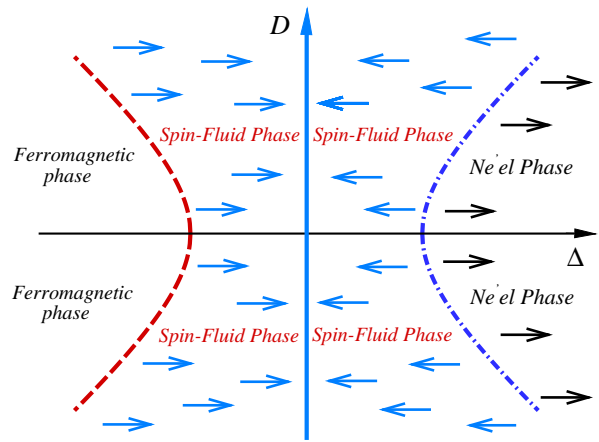}
\caption{(color online) Phase diagram of the XXZ model with DM
interaction. The long dashed line is the critical line which
separates SF-N\'{e}el phases and is characterized by
$\Delta_{c}=(1+D^{2})^{1/2}$. The dashed dot line shows the level
crossing  which separate SF-F phases is given by
$\Delta=-(1+D^{2})^{1/2}$ . Arrows show the running of coupling
constant under RG iteration.} \label{fig2}
\end{center}
\end{figure}
%%%%%%%%%%%%%%%%%%%%%%%%%%%%%%%%%%%%%%%%%%%%%%%%%%%%%%%

\subsection{Region (A): $e_{0}$ is the ground state.}

In this region the effective Hamiltonian in the first order
correction is similar to the initial one, i.e,

\bea \no
H^{eff}=\frac{J'}{4}\sum_{i}^{N}\Big[\sigma_{i}^{x}\sigma_{i+1}^{x}+\sigma_{i}^{y}\sigma_{i+1}^{y}+
{\Delta'}\sigma_{i}^{z}\sigma_{i+1}^{z}\\
+D'(\sigma_{i}^{x}\sigma_{i+1}^{y}-\sigma_{i}^{y}\sigma_{i+1}^{x})\Big]
\eea

where $J'$, $\Delta'$ and $D'$ are the renormalized coupling constants. The
new renormalized coupling constants are found to be functions of
the original ones given by the following equations,

\bea
\label{rgeq}
J'=J(\frac{2}{q})^{2}(1+D^{2}),~~D'=D\\
\Delta'=\frac{\Delta}{1+D^{2}}(\frac{\Delta+q}{4})^{2}.
\no
\eea

\subsection{Region (B): $e_{3}$ is the ground state.}

In this region the effective Hamiltonian to the first
order corrections leads to the ferromagnetic Ising model
\bea
\no H^{eff}=\frac{1}{4}\left[\Delta' \sum_{i}^{N/3}
{\sigma}_{i}^{z}{\sigma}_{i+1}^{z} \right],
\eea
where
\bea
\no
\Delta'&=&J\Delta,~~J>0,~~\Delta<0.
\eea

%%%%%%%%%%%%%%%%%%%%%%%%%%%%%%%%%%%%%%%%%%%%%%%%%%%%%%%%%%%%%%%%%
\section{Phase Diagram}
\subsection{Region (A)}

For simplicity we have separated this region into positive
anisotropy and negative anisotropy sectors.

\begin{itemize}
    \item $\Delta>0$

In the positive anisotropy sector the RG equations show
that the $J$ coupling, representing the energy scale, approaching
zero by iterating RG procedure. Thus, at the zero temperature,the quantum phase transition
is the result of competition between the anisotropy ($\Delta$) and the DM coupling constant ($D$).
In the region of planar anisotropy $0<\Delta<1$, the symmetric
interactions ($D=0$) is known not to support any kind of long
range order and the ground state is the so called spin-fluid (SF)
state. Increasing the amount of anisotropy is necessary to
stabilize the spin alignment. For $\Delta>1$ the ground state is
the N\'{e}el ordered state. In the case of $D\neq0$, the
anisotropy constant ($\Delta$) and antisymmetric (DM) coupling are in
competition with each other. The latter thus destroys the ordering
tendency of the former and defers creating of N\'{e}el order. Our
RG equations show that the phase boundary between the SF and
N\'{e}el phases which depends on the DM coupling is
$\Delta_{c}=\sqrt{1+D^{2}}$ (see Fig.(\ref{fig2})) which agrees with the phase boundary
reported in Ref.\onlinecite{Alcaraz}.
This critical line coincide with boundary line which obtained by
classical approximation (see appendix C). The RG equations (Eq.(\ref{rgeq}))
express the DM coupling dose not flow under RG transformations, and the
anisotropy coupling goes to zero ($\Delta\rightarrow0$) in SF
phase while it scales to infinity ($\Delta\rightarrow\infty$) in the N\'{e}el
phase.

We have linearized the RG flow at the critical line $\Delta_{c}=\sqrt{1+D^{2}}$
and found one relevant and one marginal directions. The
eigenvalues of the matrix of linearized flow are
$\lambda_1=\frac{5}{3}$, $\lambda_2=1$. The corresponding
eigenvectors in the $|\Delta, D\rangle$ coordinates are
$|\lambda_1\rangle=|1,0\rangle$,
$|\lambda_2\rangle=|\frac{D}{\sqrt{1+D^{2}}}, 1\rangle$. The
marginal direction corresponds to the tangent line of the critical line
and the relevant direction shows the direction of anisotropy's
flow (Fig.(\ref{fig2})).

%%%%%%%%%%%%%%%%%%%%%  Fig.3   %%%%%%%%%%%%%%%%%%%%%%%
\begin{figure}
\begin{center}
\includegraphics[width=8cm]{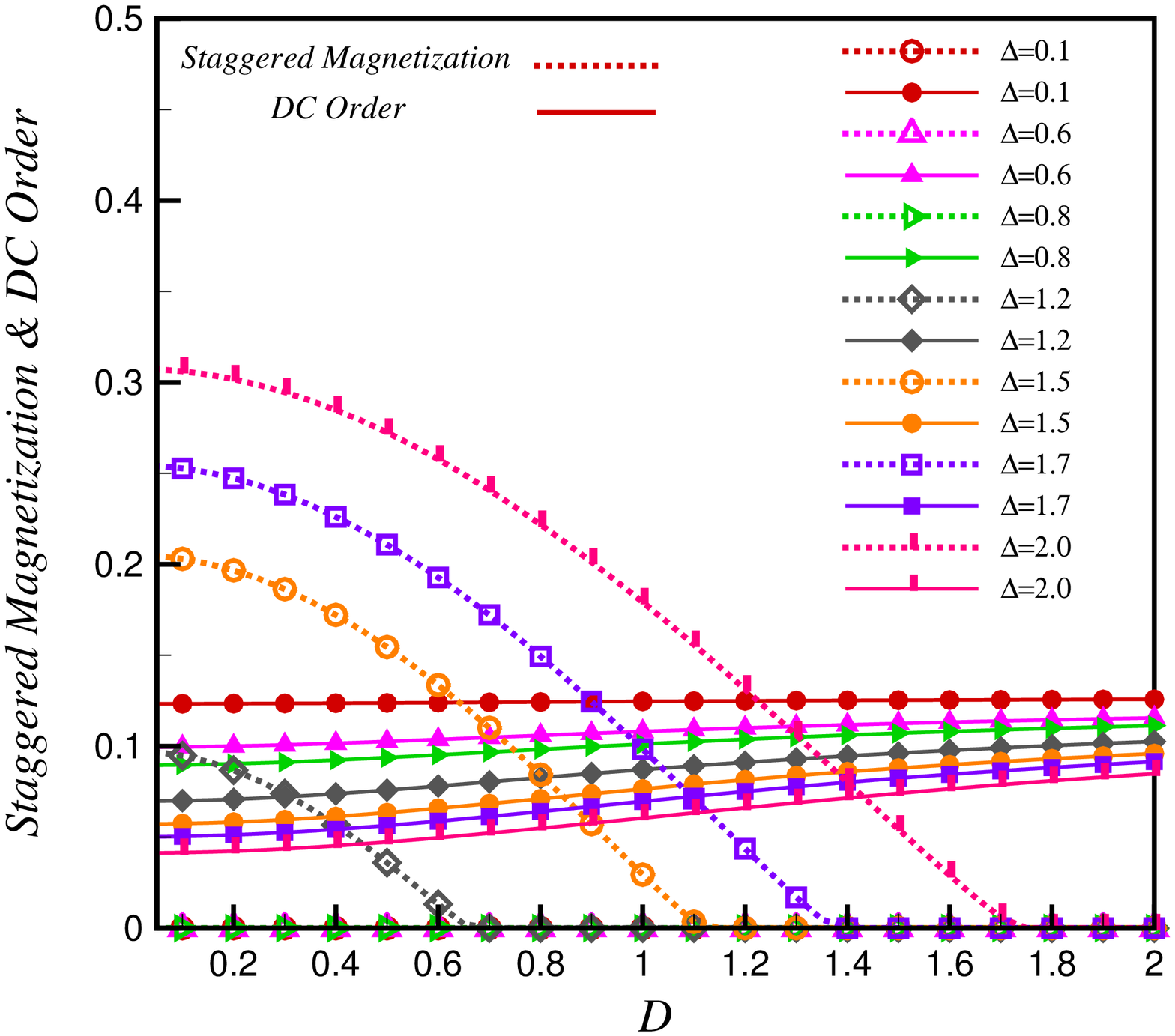}
\caption{(color online) DC order (solid lines) and Staggered
Magnetization (dotted lines) versus D.} \label{fig3}
\end{center}
\end{figure}
%%%%%%%%%%%%%%%%%%%%%%%%%%%%%%%%%%%%%%%%%%%%%%%%%%%%%%%

%%%%%%%%%%%%%%%%%%%%%  Fig.4   %%%%%%%%%%%%%%%%%%%%%%%
\begin{figure}
\begin{center}
\includegraphics[width=8cm]{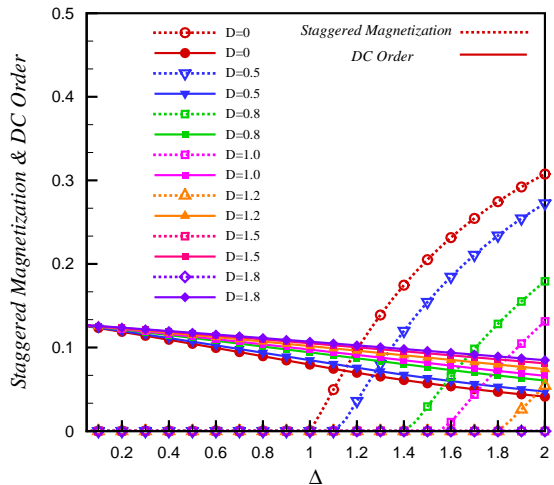}
\caption{(color online) DC order (solid lines) and Staggered
Magnetization (dotted lines) versus $\Delta$.} \label{fig4}
\end{center}
\end{figure}

%%%%%%%%%%%%%%%%%%%%%%%%%%%%%%%%%%%%%%%%%%%%%%%%%%%%%%%

However we have found the boundary of the SF-N\'{e}el transition
by calculating the staggered magnetization $S_{M}$ (see appendix
B) in the $z$-direction as an order parameter (Fig.(\ref{fig3})
and Fig.(\ref{fig4})),

\be \label{sm} S_M=\frac{1}{N}\sum_{i=1}^{N} (\frac{(-1)^i}{2})
\langle \sigma_i^z \rangle. \ee

$S_M$ is zero in the SF phase and has a nonzero value in the
N\'{e}el phase. Thus the staggered magnetization is the proper
order parameter to represent the SF-N\'{e}el transition. We have
plotted $S_{M}$ versus $D$ and versus $\Delta$ in Fig.(\ref{fig3})
and Fig.(\ref{fig4}), respectively. In Fig.(\ref{fig3}) it is obvious that the
staggered magnetization goes to zero continuously at the critical
value of $D_{c}$ which shows the destruction of N\'{e}el order.
The critical value $D_{c}$ where the staggered magnetization
vanishes above it, increases by increasing of anisotropy, this
means SF-N\'{e}el transition point ($\Delta_{c}$) depends on the
DM interaction. It is seen in  Fig.(\ref{fig4})  that the
staggered magnetization is zero for $\Delta<\Delta_{c}$ (SF phase)
and has a nonzero value for $\Delta>\Delta_{c}$ (N\'{e}el phase),
while enhancing of DM coupling defer the creation of N\'{e}el order.

Moreover, to study the influence of DM coupling we have  calculated
the chiral order \cite{Kawamura,Kaburagi}($C_{h}$) in the $z$
direction  as the helical magnetization in one
dimension\cite{Jafari2} which is created by DM interaction.
\bea \no
C_{h}=\frac{1}{N}\sum_{i=1}^{N}\frac{1}{4}\langle(\sigma_{i}^{x}\sigma_{i+1}^{y}-\sigma_{i}^{y}\sigma_{i+1}^{x})\rangle.
\eea
Unfortunately the chiral order is not a self similar operator
under RG transformations. In fact an $XX$ term of
Hamiltonian
($\sigma_{i}^{x}\sigma_{i+1}^{x}+\sigma_{i}^{y}\sigma_{i+1}^{y}$)
shows up in the renormalized chiral order under RG. Thus, we have to calculate
the sum of chiral and $XX$ term which we have represented it by
$DC$ in the following equation,

\bea \no
DC=\frac{1}{N}\sum_{i=1}^{N}\frac{1}{4}\langle(\sigma_{i}^{x}\sigma_{i+1}^{x}+\sigma_{i}^{y}\sigma_{i+1}^{y})\\
\label{DC}
+D(\sigma_{i}^{x}\sigma_{i+1}^{y}-\sigma_{i}^{y}\sigma_{i+1}^{x})\rangle.
\eea

In Fig.(\ref{fig3}) and Fig.(\ref{fig4}) the DC order has plotted
versus $D$ and $\Delta$ respectively, for different values of
$\Delta$ and $D$. The figures manifest that the $DC$ order
enhances by increasing of DM coupling and reduces with increasing
of anisotropy parameter.

We have also calculated the critical exponents at the critical
line $(\Delta_{c}=\sqrt{1+D^{2}})$. In this respect, we have
obtained the dynamical exponent, the exponent of order parameter
and the diverging exponent of the correlation length. This
corresponds to reach the critical point from the N\'{e}el phase
by approaching $\Delta\rightarrow\Delta_{c} $. The dynamical
exponent is $z \simeq 0.73$, the staggered magnetization goes to
zero like $S_M \sim |\Delta -\Delta_{c}|^{\beta}$ with
$\beta\simeq1.15$. The correlation length diverges $\xi \sim
|\Delta -\Delta_{c}|^{-\nu}$ with exponent $\nu\simeq2.15$. The
remarkable result of these exponents is the independence of their
values on the D value and equality of them with the correspondening ones
in the $XXZ$ model. The detail of this calculation is
similar to what is presented in Ref.\cite{Langari}.

    \item $\Delta<0$

In this sector, the effective Hamiltonian is similar to the
positive anisotropy case with the same coupling constants. For
$-\sqrt{1+D^{2}}<\Delta<0$ the ground state is the spin-fluid
phase and decreasing the anisotropy causes the ground state of the
three site Hamiltonian changes by level crossing at
$\Delta=-\sqrt{1+D^{2}}$ where the RG equations should be reconstructed.
However, the remarkable result is that the level crossing
line which got by three site Hamiltonian coincides the critical
line of this model in the thermodynamic system\cite{Alcaraz}. The RG
equations express the DM coupling dose not flow under  RG
transformations, and the anisotropy coupling goes to zero
($\Delta\rightarrow0$). Thus the $\Delta=0$ line is the position of
stable fixed points where the RG flow is freezed.

\end{itemize}

\subsection{Region (B)}
As we pointed out in sec.\ref{sec2}-B the original Hamiltonian is mapped to the
ferromagnetic Ising model. Ising model remains unchanged under
RG as the stable fixed point and its properties are well known. We call this region as the
ferromagnetic Ising phase.

%%%%%%%%%%%%%%%%%%%%%%%%%%%%%%%%%%%%%%%%%%%%%%%%%%%%%%%

\section{Summery and conclusions \label{conclusion}}

We have applied the RG transformation to obtain the phase diagram,
staggered magnetization and helical magnetization of $XXZ$ model
with DM interaction. In the positive anisotropy region, tuning the
anisotropy coupling makes the system to fall into different
phases, i.e N\'{e}el phase with nonzero order parameter and
spin-fluid one with vanishing order parameter as characterized by the
staggered magnetization. The RG equations state that the system
has fixed points at $\Delta=0$, $\Delta=\infty$ and
$\Delta_{c}=\sqrt{1+D^{2}}$. The fixed points $\Delta=0$ and
$\Delta=\infty$ are attractive and correspond to the Spin-Fluid
and N\'{e}el phases, respectively. The fixed points at
$\Delta_{c}=\sqrt{1+D^{2}}$ are repulsive and correspond to the
critical points of this model, in the other word it is the
critical line of this Hamiltonian. However, in the negative anisotropy
region, the level crossing line $\Delta=-\sqrt{1+D^{2}}$ which is
obtained by three site block Hamiltonian eigenvalues, is the
critical line of infinite size system and separates the
ferromagnetic and spin-fluid phases. Unfortunately we can not
calculate the chiral order explicitly by RG method. To survey the
influence of DM interaction and helical magnetization, the numerical
Lanczos computation is in progress.

%%%%%%%%%%%%%%%%%%%%%%%%%%%%%%%%%%%%%%%%%%%%%%%%%%%%%%%%%%%%%%%%%
\begin{acknowledgments}
The authors would like to thank J. Abouie, M. Kargarian and M. Siahatgar
for fruitful discussions.
This work was supported in part by the Center of Excellence in
Complex Systems and Condensed Matter (www.cscm.ir).
\end{acknowledgments}
%%%%%%%%%%%%%%%%%%%%%%%%%%%%%%%%%%%%%%%%%%%%%%%%%%%%%%%%%%%%%%%%

\section{Appendix \label{Appendix}}

\subsection{The block Hamiltonian of three sites, its eigenvectors and
eigenvalues}

%%%%%%%%%%%%%%%%%%%%%  Fig.5   %%%%%%%%%%%%%%%%%%%%%%%
\begin{figure}
\begin{center}
\includegraphics[width=8cm]{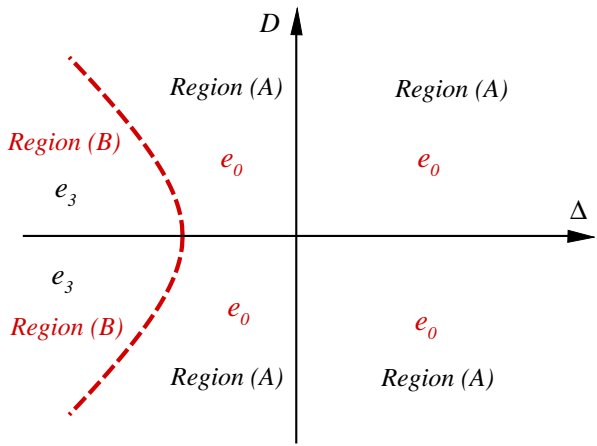}
\caption{(color online) The ground state eigenvalues as a function of
anisotropy and DM coupling. The thick long dashed line which shows the
border line of region (A) and region (B) are given by $\Delta=-\sqrt{1+D^{2}}$.} \label{fig5}
\end{center}
\end{figure}

%%%%%%%%%%%%%%%%%%%%%%%%%%%%%%%%%%%%%%%%%%%%%%%%%%%%%%%

We have considered the three-site block (Fig.(\ref{fig1})) with the
following Hamiltonian

%\begin{widetext}
\bea \no
h_{I}^{B}=\frac{J}{4}\Big[(\sigma_{1,I}^{x}\sigma_{2,I}^{x}+\sigma_{2,I}^{x}\sigma_{3,I}^{x}+
\sigma_{1,I}^{y}\sigma_{2,I}^{y}+\sigma_{2,I}^{y}\sigma_{3,I}^{y})\\
\no
+\Delta(\sigma_{1,I}^{z}\sigma_{2,I}^{z}+\sigma_{2,I}^{z}\sigma_{3,I}^{z})~~~~~~~~~~~~~~~~~~~~~~~~~~~~~\\
\no
+D(\sigma_{1,I}^{x}\sigma_{2,I}^{y}-\sigma_{1,I}^{y}\sigma_{2,I}^{x}+
\sigma_{2,I}^{x}\sigma_{3,I}^{y}-\sigma_{2,I}^{y}\sigma_{3,I}^{x})\Big]
\eea
%\end{widetext}

The inter-block ($H^{BB})$ and intra-block ($H^{B}$) Hamiltonian for
the three sites decomposition are
%\begin{widetext}
\bea \no
H^{BB}=\frac{J}{4}\sum_{I}^{N/3}\left[(\sigma_{3,I}^{x}\sigma_{1,I+1}^{x}+
\sigma_{3,I}^{y}\sigma_{1,I+1}^{y}+\Delta\sigma_{3,I}^{z}\sigma_{1,I+1}^{z})\right.\\
\no
\left.+D(\sigma_{3,I}^{x}\sigma_{1,I+1}^{y}-\sigma_{3,I}^{y}\sigma_{1,I+1}^{x})\right]\\,
\no
H^{B}=\frac{J}{4}\sum_{I}^{N/3}\left[(\sigma_{1,I}^{x}\sigma_{2,I}^{x}+\sigma_{2,I}^{x}\sigma_{3,I}^{x}+
\sigma_{1,I}^{y}\sigma_{2,I}^{y}+\sigma_{2,I}^{y}\sigma_{3,I}^{y})\right.\\
\no
\left.+\Delta(\sigma_{1,I}^{z}\sigma_{2,I}^{z}+\sigma_{2,I}^{z}\sigma_{3,I}^{z})\right.~~~~~~~~~~~~~~~~~~~~~~~~~~~~~\\
\no
\left.+D(\sigma_{1,I}^{x}\sigma_{2,I}^{y}-\sigma_{1,I}^{y}\sigma_{2,I}^{x}+
\sigma_{2,I}^{x}\sigma_{3,I}^{y}-\sigma_{2,I}^{y}\sigma_{3,I}^{x})\right].
\eea
%\end{widetext}

where $\sigma_{j,I}^{\alpha}$ refers to the $\alpha$-component of
the Pauli matrix at site $j$ of the block labeled by $I$. The exact
treatment of this Hamiltonian leads to four distinct eigenvalues
which are doubly degenerate. The ground, first, second and third
excited state energies have the following expressions in terms of
the coupling constants.

\begin{widetext}
\begin{eqnarray}
|\psi_{0}\rangle=\frac{1}{\sqrt{2q(q+\Delta)(1+D^{2})}}\Big[2(D^{2}+1)|\downarrow\downarrow\uparrow\rangle-
(1-iD)(\Delta+q)|\downarrow\uparrow\downarrow\rangle-2[2iD+(D^{2}-1)]|\uparrow\downarrow\downarrow\rangle\Big],\\
\no
|\psi_{0}'\rangle=\frac{1}{\sqrt{2q(q+\Delta)(1+D^{2})}}\Big[2(D^{2}+1)|\downarrow\uparrow\uparrow\rangle-
(1-iD)(\Delta+q)|\uparrow\downarrow\uparrow\rangle-2[2iD+(D^{2}-1)]|\uparrow\uparrow\downarrow\rangle)\Big],\\
\label{eqA1}
e_{0}=-\frac{J}{4}(\Delta+q),~~~~~~~~~~~~~~~~~~~~~~~~~~~~~~~~~~~~~~~~~~~~~~~\\
\no
|\psi_{1}\rangle=\frac{1}{\sqrt{2q(q-\Delta)(1+D^{2})}}\Big[2(D^{2}+1)|\downarrow\downarrow\uparrow\rangle-
(1-iD)(\Delta-q)|\downarrow\uparrow\downarrow\rangle-2[2iD+(D^{2}-1)]|\uparrow\downarrow\downarrow\rangle\Big],\\
\no
|\psi_{1}'\rangle=\frac{1}{\sqrt{2q(q-\Delta)(1+D^{2})}}\Big[2(D^{2}+1)|\downarrow\uparrow\uparrow\rangle-
(1-iD)(\Delta-q)|\uparrow\downarrow\uparrow\rangle-2[2iD+(D^{2}-1)]|\uparrow\uparrow\downarrow\rangle)\Big],\\
\no
e_{0}=-\frac{J}{4}(\Delta-q),~~~~~~~~~~~~~~~~~~~~~~~~~~~~~~~~~~~~~~~~~~~~~~~\\
\no
|\psi_{2}\rangle=\frac{1}{\sqrt{2}(1+D^{2})}\Big[[2iD+(D^{2}-1)]|\uparrow\downarrow\downarrow\rangle+
(D^{2}-1)|\downarrow\downarrow\uparrow\rangle\Big],~~~~~~~~~~~~~~~~~~~~~~~~\\
\no
|\psi_{2}'\rangle=\frac{1}{\sqrt{2}(1+D^{2})}\Big[[2iD+(D^{2}-1)]|\uparrow\uparrow\downarrow\rangle+
(D^{2}-1)|\downarrow\uparrow\uparrow\rangle\Big],~~~~~~~~~~~~~~~~~~~~~~~~\\
e_{2}=0,~~~~~~~~~~~~~~~~~~~~~~~~~~~~~~~~~~~~~~~~~~~~~~~~~~~~~~~\\
\no |\psi_{3}\rangle=|\uparrow\uparrow\uparrow\rangle~~~,~~~
|\psi_{3}'\rangle=|\downarrow\downarrow\downarrow\rangle,~~~~~~~~~~~~~~~~~~~~~~~~~~~~~~~~~~~~~~~~~~~~\\
\no
e_{3}=\frac{J}{2}(\Delta),~~~~~~~~~~~~~~~~~~~~~~~~~~~~~~~~~~~~~~~~~~~~~~~~~~~~~~~~
\end{eqnarray}
\end{widetext}

where $q$ is $q=\sqrt{\Delta^{2}+8(1+D^{2})}$.

$|\uparrow\rangle$ and $|\downarrow\rangle$ are the eigenstates of
$\sigma^{z}$. In Fig.(\ref{fig5}) we have presented the different regions where
the specified state is the ground state of the block Hamiltonian.
In the region (A) the projection operator is
\bea \no
P_{0}=|\Uparrow\rangle\langle\psi_{0}|+|\Downarrow\rangle\langle\psi_{0}'|.
\eea
The Pauli matrices in the effective Hilbert space have the following
transformations
\begin{widetext}
\bea \no
P_{0}^{I}\sigma_{1,I}^{x}P_{0}^{I}=-\frac{2}{q}({\sigma'}_{I}^{x}+D{\sigma'}_{I}^{y})~~~,~~~
P_{0}^{I}\sigma_{2,I}^{x}P_{0}^{I}=\frac{4(D^{2}+1)}{q(\Delta+q)}{\sigma'}_{I}^{x}~~~,~~~
P_{0}^{I}\sigma_{3,I}^{x}P_{0}^{I}=-\frac{2}{q}({\sigma'}_{I}^{x}-D{\sigma'}_{I}^{y})\\
\no
P_{0}^{I}\sigma_{1,I}^{y}P_{0}^{I}=-\frac{2}{q}(D{\sigma'}_{I}^{x}-{\sigma'}_{I}^{y})~~~,~~~
P_{0}^{I}\sigma_{2,I}^{y}P_{0}^{I}=-\frac{4(D^{2}+1)}{q(\Delta+q)}{\sigma'}_{I}^{y}~~~,~~~
P_{0}^{I}\sigma_{3,I}^{y}P_{0}^{I}=\frac{2}{q}(D{\sigma'}_{I}^{x}+{\sigma'}_{I}^{y})\\
\no
P_{0}^{I}\sigma_{1,I}^{z}P_{0}^{I}=P_{0}^{I}\sigma_{3,I}^{z}P_{0}^{I}=-\frac{\Delta+q}{2q}{\sigma'}_{I}^{z}~~~,~~~
P_{0}^{I}\sigma_{2,I}^{z}P_{0}^{I}=\frac{\Delta}{q}{\sigma'}_{I}^{z}~~~~~~~~~~~~~~~~~~~~~~~
\eea
\end{widetext}

In the region (B) ($\Delta<-\sqrt{1+D_{2}}$) the projection operator is
\bea \no
P_{0}=|\Uparrow\rangle\langle\psi_{3}|+|\Downarrow\rangle\langle\psi_{3}'|.
\eea
and the Pauli matrices in the effective Hilbert space have the following
transformations

\bea \no
P_{0}^{I}\sigma_{1,I}^{x}P_{0}^{I}=0~~~,~~~
P_{0}^{I}\sigma_{2,I}^{x}P_{0}^{I}=0~~~,~~~
P_{0}^{I}\sigma_{3,I}^{x}P_{0}^{I}=0\\
\no
P_{0}^{I}\sigma_{1,I}^{y}P_{0}^{I}=0~~~,~~~
P_{0}^{I}\sigma_{2,I}^{y}P_{0}^{I}=0~~~,~~~
P_{0}^{I}\sigma_{3,I}^{y}P_{0}^{I}=0\\
\no
P_{0}^{I}\sigma_{1,I}^{z}P_{0}^{I}=P_{0}^{I}\sigma_{3,I}^{z}P_{0}^{I}=P_{0}^{I}\sigma_{2,I}^{z}P_{0}^{I}=
{\sigma'}_{I}^{z}.
\eea

\subsection{Order Parameter and Chiral Order}

\subsubsection{Staggered magnetization}

Generally, any correlation function can be calculated in the QRG
scheme. In this approach, the correlation function at each iteration of
RG is connected to its value after an RG iteration. This will be
continued to reach a controllable fixed point where we can obtain
the value of the correlation function. The staggered magnetization
in $\alpha$ direction can be written
\bea \label{eqB1} S_{M}=\frac{1}{N}\sum_{i}^{N}\langle
O|\frac{(-1)^i}{2}\sigma_{i}^{\alpha}|O\rangle, \eea
where $\sigma_{i}^{\alpha}$ is the Pauli matrix in the $i$th site
and $|O\rangle$ is the ground state of chain. The ground state of
the renormalized chain is related to the ground state of the
original one by the transformation, $P_{0}|O'\rangle=|O\rangle$.
\bea \no S_{M}=\frac{1}{N}\sum_{i}^{N}\langle
O'|P_{0}(\frac{(-1)^i}{2}\sigma_{i}^{\alpha})P_{0}|O'\rangle. \eea
This leads to the staggered configuration in the renormalized chain.
The staggered magnetization in $z$ direction is obtained

%\begin{widetext}
\begin{eqnarray}
\label{eqB2}
\no S_{M}^{0}&=&\frac{1}{N}\sum_{i=1}^{N}\langle
0|\frac{(-1)^i}{2}\sigma_{i}^{z}|0\rangle\\
\no
&=&\frac{1}{6}\frac{1}{\frac{N}{3}}\sum_{I=1}^{N/3}\Big[\langle
0'|P_{0}^{I}(-\sigma_{1,I}^{z}+
\sigma_{2,I}^{z}-\sigma_{3,I}^{z})P_{0}^{I}|0'\rangle\\
\no
&-&\langle0'|P_{0}^{I+1}(-\sigma_{1,I+1}^{z}+\sigma_{2,I+1}^{z}-\sigma_{3,I+1}^{z})P_{0}^{I+1}|0'\rangle\Big]\\
&=&-\frac{2\Delta+q}{3q}\frac{1}{\frac{N}{3}}\sum_{I=1}^{N/3}\langle
0'|\frac{(-1)^i}{2}\sigma_{I}^{z}|0'\rangle=-\frac{\gamma^{0}}{3}S_{M}^{1},
\end{eqnarray}
%\end{widetext}
\\
where $S_{M}^{(n)}$ is the staggered magnetization at the $n$th
step of QRG and $\gamma^{(0)}$ is defined by
$\gamma^{0}=(2\Delta+q)/q$.

This process will be iterated many times by replacing
$\gamma^{(0)}$ with $\gamma^{(n)}$. The expression for
$\gamma^{(n)}$ is similar to $\gamma^{(0)}$ where the coupling
constants should be replaced by the renormalized ones at the
corresponding RG iteration ($n$). The result of this calculation
has been presented in Fig.(\ref{fig3}) and Fig.(\ref{fig4}).

\subsubsection{Chiral Order}

The chiral order which is the proper function to detect the
helical magnetization in the systems can be written

\be \label{Ch}
C_{h}=\frac{1}{N}\sum_{i=1}^{N}\frac{1}{4}\langle(\sigma_{i}^{x}\sigma_{i+1}^{y}-\sigma_{i}^{y}\sigma_{i+1}^{x})\rangle.
\ee

As we mentioned in the section III, the $XX$ term of the Hamiltonian
shows up to the chiral order under RG. The $XX$ term order is
written

\be \label{Dim}
D_{XX}=\frac{1}{N}\sum_{i=1}^{N}\frac{1}{4}\langle(\sigma_{i}^{x}\sigma_{i+1}^{x}+\sigma_{i}^{y}\sigma_{i+1}^{y})\rangle.
\ee

In this case the calculating of the chiral order being elaborate,
because of the unknown effect of the $XX$ term on the ground
state of system at fixed point ($\Delta=\infty$). To
simplification of the calculation, we transform the $XXZ$ with DM
interaction Hamiltonian (Eq.(\ref{eq1})) to the Ising model with
DM interaction\cite{Jafari2} (IDM) by implement a non-local transformation
which shows a $n\varphi$
rotation about the $z$ axis at site $n$ where
$\varphi=\arctan(-\frac{1}{D})$. We have calculated the
chiral order (Eq.(\ref{Ch})) of IDM in Ref.[\cite{Jafari2}]. By
implement the inverse of transformation, the chiral order in IDM
model transforms to the DC order where introduced in
Eq.(\ref{DC}). The DC order has been plotted in Fig.(\ref{fig3})
and Fig.(\ref{fig4}) versus $D$ and $\Delta$.

\subsection{Classical Approximation}\label{classical}

In the classical approximation the spins are considered as
classical vectors which form the spiral structure with a pitch
angle $\varphi$ between neighboring spins and canted angle
$\theta$

\bea \no \sigma^{x}_{n}=\cos(n\varphi)\sin\theta~~,
\sigma^{y}_{n}=\sin(n\varphi)\sin\theta~~,
\sigma^{z}_{n}=\cos\theta,\eea

The classical energy per site for the $XXZ$ with DM interaction
Hamiltonian (Eq.(\ref{eq1})) is

 \bea \no
\frac{E_{cl}}{N}=\frac{J}{4}[(\cos\varphi+D\sin\varphi)\sin^{2}\theta+\Delta\cos^{2}\theta].
\eea

\begin{figure}
\begin{center}
\includegraphics[width=8cm]{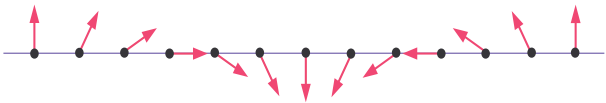}
\caption{(color online) A classical picture of spin orientation in
the $xy$ plain where the angle between neibouring spins depend on
the D value.} \label{figCh}
\end{center}
\end{figure}

The minimization of classical energy with respect to the angles
$\varphi$ and $\theta$ shows that there are two different regions.
(I) $\Delta>\sqrt{1+D^{2}}$, the minimum of energy is obtained by
arbitrary $\theta$ and $\varphi=\arctan(D)$ which show the spins
projection on $z$ axis is nonzero and spins have the helical
structure (see Fig.\ref{figCh}) in the $xy$ plain. In this region
the minimum classical energy is \bea \label{eqC1}
\frac{E_{cl}^{I}}{N}=\frac{J}{4}(\sqrt{1+D^{2}}\sin^{2}\theta+\Delta\cos^{2}\theta).
\eea

(II) $\Delta<\sqrt{1+D^{2}}$, the energy is minimized by
$\Delta=\cos\varphi+D\sin\varphi$ and arbitrary $\theta$ or
arbitrary $\varphi$ and $\theta=\frac{\pi}{2}$, which correspond
respectively to the configurations with nonzero value of spins
projection on $z$-axis with helical structure of spins projection
in the $xy$-plain and disorder configuration. In this region the
minimum classical energy is

\bea \label{eqC2}
\frac{E_{cl}^{II}}{N}=\frac{J}{4}(\cos\varphi+D\sin\varphi)=\frac{J}{4}\Delta.
\eea

One can see from Eq.(\ref{eqC1}) and Eq.(\ref{eqC2}) that the
transition between phase (I) and (II) takes place at
$\Delta=\sqrt{1+D^{2}}$.

%%%%%%%%%%%%%%%%%%%%%  Fig.Ch   %%%%%%%%%%%%%%%%%%%%%%%
%%%%%%%%%%%%%%%%%%%%%%%%%%%%%%%%%%%%%%%%%%%%%%%%%%%%%%%

\section{Canonical Transformation}

The Hamiltonian of $XXZ$ model with DM interaction (Eq.(\ref{eq1})) has the global $U(1)\times Z_{2}$ symmetry. This Hamiltonian is mapped
to the well known $XXZ$ chain via a canonical transformation\cite{Aristov,Alcaraz},

\bea \label{eqD2}\no
U=\sum_{j=1}^{N}\alpha_{j}\sigma_{j}^{z}~~~&,&~~~\alpha_{j}=\sum^{j-1}_{m=1}
m \tan^{-1}(D), \\
 \no \tilde{\sigma}^{\pm}_{j}=e^{-iU}\sigma^{\pm}_{j}e^{iU}&,&~~~~~
\tilde{\sigma}^{z}_{j}=\sigma^z,\\
 \tilde{H}&=&e^{-iU} H e^{iU},
\eea

which gives

\be
\label{eqD3}
 \tilde{H}=\frac{J\sqrt{1+D^2}}{4}\Big[\sum_i\tilde{\sigma}^{x}_{i} \tilde{\sigma}^{x}_{i+1}+
\tilde{\sigma}^{y}_{i} \tilde{\sigma}^{y}_{i+1}+
(\frac{\Delta}{\sqrt{1+D^2}})\tilde{\sigma}^{z}_{i} \tilde{\sigma}^{z}_{i+1}\Big].
\ee
The $U(1)\times Z_{2}$ symmetry of initial Hamiltonian survive in the transformed Hamiltonian too, but at $\Delta_{c}=\pm\sqrt{1+D^{2}}$ the $U(1)\times Z_{2}$ symmetry breaks to the local $SU(2)$ symmetry.

\end{document}